\documentclass[sigconf,nonacm]{acmart}

\usepackage{booktabs}
\usepackage{enumitem}
\usepackage{microtype}
\usepackage{cleveref}

\usepackage{tikz}
\usetikzlibrary{arrows.meta, positioning, fit, shapes.geometric}

\usepackage{url}

\title{Minions: A Dynamic and Distributed Virtual GPU Cloud Built on Idle Apple Silicon Devices}

\author{Jianhua Li}
\affiliation{%
  \institution{Apple}
  \country{USA}
}
\email{jianhua_li3@apple.com}

\author{Yun Xing}
\affiliation{%
  \institution{Apple}
  \country{USA}
}
\email{yun_xing@apple.com}

\author{Girisha De Silva}
\affiliation{%
  \institution{Apple}
  \country{USA}
}
\email{girisha_desilva@apple.com}

\author{Ryan Flint}
\affiliation{%
  \institution{Apple}
  \country{USA}
}
\email{rflint@apple.com}

\author{Kai Yuan}
\affiliation{%
  \institution{Apple}
  \country{USA}
}
\email{kai_yuan@apple.com}

\author{Sean Suchter}
\affiliation{%
  \institution{Apple}
  \country{USA}
}
\email{ssuchter@apple.com}

\begin{abstract}
The increasing demand for GPU-accelerated machine learning inference has exposed limitations in centralized GPU infrastructures, including high cost, limited elasticity, and long wait times under resource constraints. At the same time, modern Apple Silicon--based macOS devices are equipped with powerful GPUs that remain largely idle during normal usage. This paper presents Minions, a dynamic and distributed virtual GPU cloud that aggregates idle Apple Silicon GPUs across macOS devices to execute large-scale machine learning inference workloads. Minions enables teams to leverage existing hardware without requiring additional GPU procurement. The system provides flexible GPU inference task execution, fault tolerance, and efficient resource utilization. Our experience shows that idle macOS devices can be transformed into a practical distributed GPU inference platform in environments where centralized GPU resources are constrained.
\end{abstract}


\begin{document}

\maketitle

\section{Introduction}
GPU-accelerated machine learning inference workloads---such as embedding generation and large language model (LLM) completion---are core components of modern information retrieval systems. In many industrial environments, these workloads depend on centralized GPU clusters that are expensive to expand and frequently oversubscribed. When GPU resources are constrained, teams may experience long queue times, reduced experimentation velocity, and increased operational cost.

In parallel, the widespread adoption of Apple Silicon has introduced a large population of macOS devices with high-performance GPUs. These devices are primarily used for interactive workloads, leaving significant GPU capacity idle for extended periods. This observation motivates the central question behind Minions: can idle Apple Silicon GPUs be dynamically aggregated into a shared, distributed inference platform without additional hardware investment?

Minions is designed as a distributed virtual GPU cloud that complements centralized GPU infrastructure. It focuses on unlocking idle compute capacity in environments where GPU resources are constrained, enabling teams to execute inference workloads immediately using existing macOS devices.

\section{System Overview and Architecture}
\subsection{System Architecture and Workflow}
Minions consists of four primary components: Minion Server, PostgreSQL Metadata Store, Minion Executor, and Minion Watchdog.

\begin{figure*}[t]
  \centering
  \includegraphics[width=\textwidth]{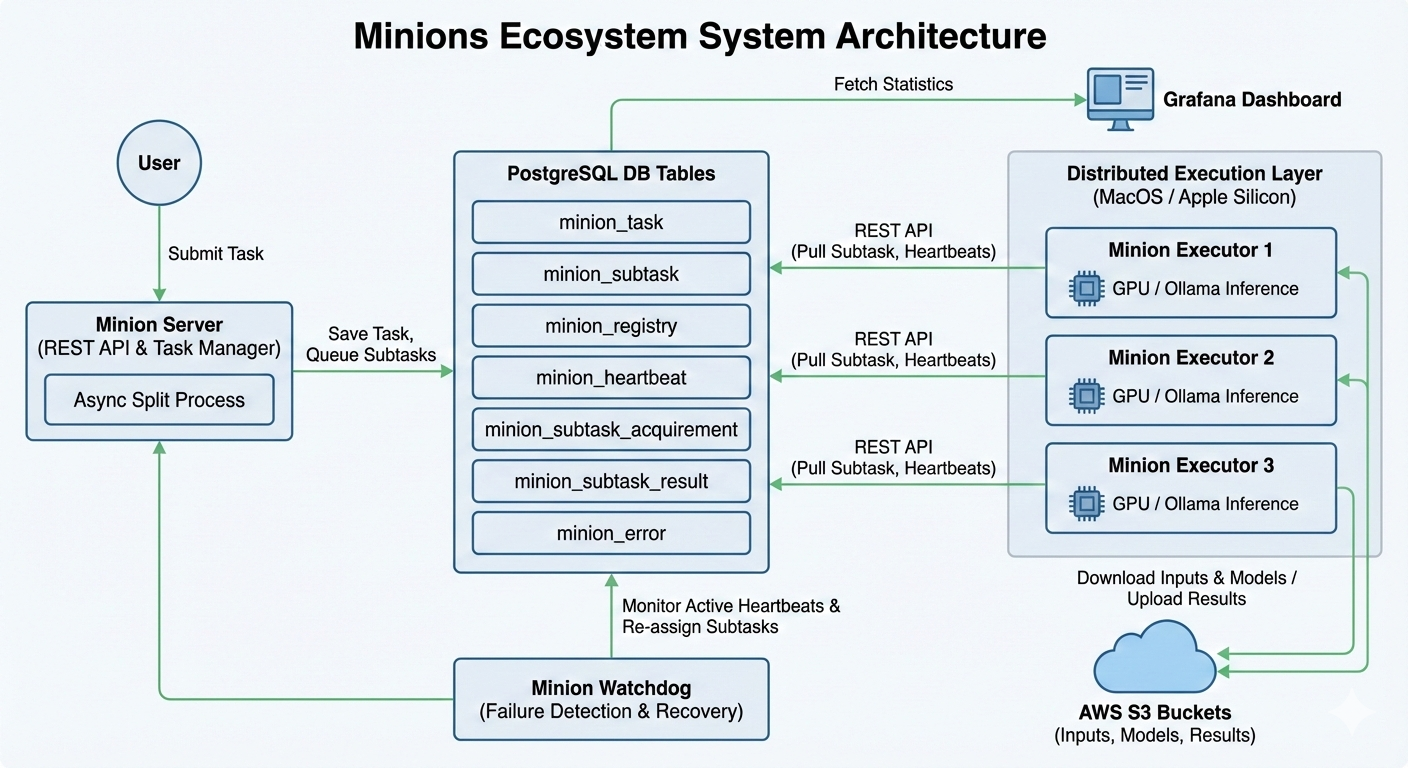}
  \caption{Minions architecture overview.}
  \label{fig:architecture}
\end{figure*}

\Cref{fig:architecture} illustrates the end-to-end workflow of GPU inference task submission, execution, and failure recovery in Minions.

Users or IR pipelines submit GPU inference tasks through the Minion Server, which validates task parameters and asynchronously partitions each task into multiple subtasks. These subtasks, along with their execution metadata, are persisted in the PostgreSQL metadata store.

Minion Executors run on macOS devices and dynamically pull eligible subtasks via REST APIs exposed by the Minion Server. This pull-based execution model accommodates the inherently intermittent availability of personal devices while enabling elastic parallelism. Executors execute subtasks locally using Apple Silicon GPUs and write inference outputs to a configured object store (e.g., Amazon S3).

Executor failures are expected in this environment due to factors such as network loss, device sleep, power-off events, etc. In such cases, the Minion Watchdog reclaims unfinished subtasks and makes them available for re-execution.

\subsection{Key Components}
\paragraph{Minion Server.}
The Minion Server serves as the central coordination layer of the system, handling all user-facing requests through a web UI and APIs, including task creation, parameter validation, task status retrieval, and result display.

Upon task submission, the Minion Server asynchronously splits a task into subtasks and enqueues them into the PostgreSQL metadata store. To protect the database and keep executors lightweight, Minion Executors never access PostgreSQL directly. Instead, the Minion Server exposes REST APIs that allow executors to acquire subtasks, update task and subtask state, write heartbeat records, and publish execution results.

\paragraph{PostgreSQL metadata store.}
Minions uses a PostgreSQL metadata store~\cite{postgresql} to maintain all task- and executor-related state. The system consists of seven tables as shown in \Cref{fig:architecture}. Separate development and production database instances are maintained to support safe end-to-end local development and testing.

\paragraph{Minion Executor.}
The Minion Executor is the primary GPU execution agent and runs on macOS devices equipped with Apple Silicon GPUs. Executors periodically fetch eligible subtasks via the Minion Server, atomically acquire execution ownership, and download subtask inputs from the configured object store. If a required language model is not already cached locally, the executor downloads and caches the model before execution.

Minions currently supports two categories of GPU inference workloads on macOS devices. For embedding generation, Minions leverages open-source transformer models from HuggingFace~\cite{huggingface}. Executors use the HuggingFace \texttt{transformers} Python API to load supported embedding models and generate vector representations locally on Apple Silicon GPUs. For large language model (LLM) completion workloads, Minions integrates with Ollama, an open-source framework for running LLMs locally on macOS devices~\cite{ollama}. Ollama is installed and executed on participating devices, enabling local inference without remote API calls. Executors interact with Ollama to perform completion requests and return results through the Minions pipeline.

After execution completes, results are uploaded to the configured output location, and execution status is reported back to the system. Throughout execution, executors send regular (five-minute interval) heartbeats to the Minion Server, which persists them into the PostgreSQL DB heartbeat table.

\paragraph{Minion Watchdog.}
The Minion Watchdog is responsible for monitoring active subtask executions and handling executor failures. Rather than scanning all registered executors, the Watchdog tracks heartbeats only for executors currently running subtasks, which improves scalability and efficiency. When an executor becomes unresponsive—due to network loss, device sleep, power-off, or intentional termination by the device user when full system resources are required—the Watchdog releases the affected subtask back to the queue, allowing other healthy executors to re-acquire and execute it.

This design allows Minions to tolerate ad hoc interruptions of individual executor machines without manual intervention, ensuring reliable task completion in the presence of ad hoc device churn.

\section{Implementation Details}
\subsection{Data Storage and I/O}
Minions uses AWS S3 as the primary storage layer for GPU inference inputs and outputs. Task input artifacts are read from S3 by executors, and inference results are written back to S3 upon completion.

Using S3 decouples compute from storage, avoids transferring large payloads through the Minion Server, and leverages industry-standard access control and security mechanisms. This design simplifies data management while supporting scalable and secure execution.

\subsection{Atomic Subtask Acquisition}
Minions adopts a pull-based execution model and intentionally avoids a centralized scheduler. Subtask claiming is implemented with an atomic acquirement record in PostgreSQL. We maintain a dedicated table, \texttt{minion\_subtask\_acquirement}, that records which executor currently owns a given subtask. The table enforces a unique constraint over the tuple \texttt{(team, project, task\_name, subtask\_id)}, which guarantees that at most one ownership row can exist for any subtask at a time.

When multiple executors contend for the same subtask, they race to insert the corresponding acquirement row (e.g., using \texttt{INSERT} with conflict handling). PostgreSQL resolves the conflict: exactly one insert succeeds, and the winning executor becomes the subtask owner; all other contenders observe the uniqueness violation (or \texttt{ON CONFLICT DO NOTHING}) and must select a different subtask. This approach provides a simple, low-overhead mutual exclusion mechanism using standard database constraints, without requiring a centralized scheduler or additional logic in the Minion Watchdog.

\subsection{Dynamic Task Binding}
By default, Minions processes inference workloads in a first-in, first-out (FIFO) order: tasks submitted earlier are executed earlier when executors become available. However, teams occasionally need to prioritize a newly submitted high-urgency task. Dynamic task binding addresses this need by allowing executors to express preferences for a specific team, project, and task name.

Preferences are configured on each Minion Executor host device. Since host devices contribute their own compute resources to the Minions ecosystem, device owners can decide which teams/projects to prioritize. In practice, a team can temporarily aggregate executors from the devices it manages and configure them to target a particular \texttt{(team, project, task\_name)} tuple.

These preferences are evaluated by the Minion Server as part of subtask selection logic and translated into database-level filters. As a result, the team's executors preferentially dequeue and execute matching subtasks, even if other tasks are ahead in the global queue. When no preferred team/project tasks are pending, executors fall back to executing tasks from the global queue (including tasks from other teams and projects).

\subsection{Observability and Monitoring}
Minions incorporates observability through Grafana-based dashboards~\cite{grafana} that provide real-time visibility into system state at a team level. All observability metrics are derived from the PostgreSQL Metadata Store, which serves as the system's source of truth.
Executor-level metrics include the number of registered, online, busy, and idle executors. Task-level metrics track active, completed, and failed tasks and the inference type distribution. These dashboards help teams understand available capacity and coordinate executor participation.

\section{Evaluation and Results}
Minions has been used to support GPU inference workloads across multiple teams. Prior to its adoption, some inference jobs experienced queue times exceeding ten hours due to limited centralized GPU availability. With Minions, tasks can often begin execution within minutes by leveraging idle macOS devices. \Cref{tab:impact} summarizes the observed operational impact.

In practice, teams frequently coordinate executor participation informally. For example, team members may send Slack messages requesting others to temporarily run the Minion Executor and target a specific project or task.

\begin{table}[t]
  \caption{Operational impact of Minions.}
  \label{tab:impact}
  \centering
  \begin{tabular}{@{}lll@{}}
    \toprule
    \textbf{Metric} & \textbf{Before Minions} & \textbf{With Minions} \\
    \midrule
    GPU queue time & Hours (up to 10+) & Minutes \\
    New GPU requests & Required & None \\
    Infrastructure cost & High & Zero \\
    Scaling effort & Manual & Organic \\
    \bottomrule
  \end{tabular}
\end{table}

A common adoption concern observed during deployment is data confidentiality when executing workloads on devices outside one's immediate organization. Even for internal workloads, perceived risk can limit participation.

To mitigate this concern, Minions supports team- or organization-scoped deployments backed by dedicated PostgreSQL databases and isolated executor pools. This approach allows teams to retain strong governance controls while benefiting from distributed execution.

\section{Conclusion and Future Work}
Minions demonstrates that idle Apple Silicon GPUs on macOS devices can be aggregated into a dynamic, distributed virtual GPU cloud for machine learning inference. The platform is particularly valuable in environments where centralized GPU resources are constrained, enabling teams to unlock substantial compute capacity at zero cost using existing hardware.

Future work will extend Minions to support distributed model training in addition to inference. More broadly, Minions points toward a novel opportunity for the machine learning community: if idle macOS devices across organizations---or potentially across the global community---can be securely federated, they could form a massive, shared GPU resource pool that benefits the broader ecosystem.

\begin{acks}
We thank Scott Guo for peer review and helpful comments, and Matt Jockers for peer review and publication coordination.
\end{acks}

\end{document}